# Technical barriers for deep closed-loop geothermal

Mark McClure

mark@resfrac.com

March 2023

Submitted to arXiv March 22, 2023

## Introduction

This is the most exciting time in my lifetime for geothermal. There are many, many innovative things happening. To name a few – promising new approaches to Enhanced Geothermal Systems, geothermal projects in sedimentary and lower enthalpy formations, new approaches for geothermal exploration, lithium extraction from produced brines, geothermal energy storage, integrations with CO2 storage and capture, and new technologies for producing energy from hot water that is coproduced with oil and gas.

However, this post is about a concept about which I remain skeptical – deep closed-loop heat exchangers (McClure, 2021). These designs are sometimes called 'Advanced Geothermal Systems,' AGS (Malek et al., 2022).

To clarify, this blog post discusses the use of closed-loop heat exchangers in *deep* geothermal wells (1000s of ft deep). This is a different application than the use of closed-loop heat exchangers in *very shallow* wells (10s of ft deep) for ground source heat pumps.

The fundamental challenge is that closed-loop heat exchanger designs rely on conduction – and sometimes free convection – to bring energy into the well. These processes are inherently much slower than forced convection, which is what drives energy transport into a conventional geothermal well. Slower energy transport means less revenue, and therefore, worse economics.

In the past few years, there have been many creative new ideas for using closed-loop heat exchangers for geothermal energy production. Usually, these proposals recognize the challenge facing closed-loop, but introduce a wrinkle intended to overcome the problem.

Despite these efforts, there are still daunting technical hurdles to widespread deployment of deep closed-loop geothermal. In this post, I go through some of the proposed designs and discuss key issues.

## Purely conductive closed-loop designs

Purely conductive closed-loop designs rely solely on heat conduction to transport energy into the wellbore. Heat conduction brings energy into the wellbore slowly, and so purely conductive closed-loop designs produces very low power per ft of wellbore (McClure, 2021; van Wees, 2021; Fowler and McClure, 2021).

To address this limitation, the proposed solution is to drill extremely long wells (Toews and Holmes, 2021).

For example, Beckers and Johnston (2022) analyzed a proposal from Toews and Holmes (2021) to drill a "7.5-km deep closed-loop geothermal system consisting of 12 laterals for a total of more than 90 km of downhole well and lateral length." The laterals branch out and reconnect from two separate vertical wellbores. A schematic of the design is shown in the figure below, reproduced from Beckers and Johnston (2022).

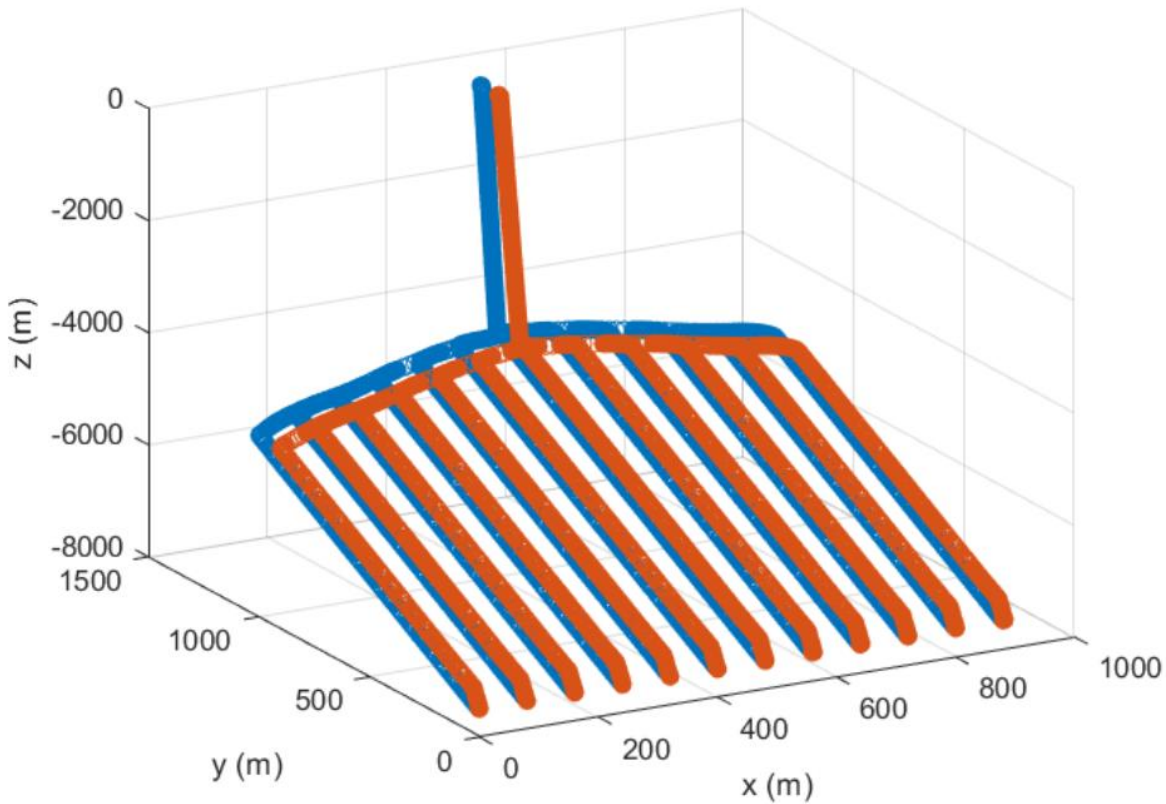

At a high geothermal gradient (60˚C/km) and medium geothermal gradient (30˚C/km) scenario, Beckers and Johnston (2022) estimated that the design would produce 8.6 MWe and 2.2 MWe, respectively.

For context, Sanyal et al. (2011) report that the typical range of conventional geothermal well production worldwide is 4-6 MWe. These are vertical or mildly deviated wells, typically drilled to a depth of 1-4 km.

Figure 7 from Lukawski et al. (2016) shows empirical statistics on geothermal well drilling as a function of depth. They report that an 'average' geothermal well drilled to 3.5 km cost $9M, with cost increasing nonlinearly with depth. Geothermal drilling costs have traditionally been higher than oil and gas drilling costs because of the challenges of temperature and geology. However, technology continues to improve over time. So, relative to the values from Lukawski et al. (2016), it's conceivable that costs could go down 2x, 3x, perhaps 4x.

Let's hypothetically suppose a 3.5 km geothermal well could be drilled for $2.5M. Given that, what would it cost to drill a pair of wells to 7.5 km, connected by 12 multilaterals with 90 km of total wellbore length?

This design would be approaching the depth of the deepest wells ever drilled in onshore North America; it would probably be a record for most laterals drilled off a single well; and it would almost certainly be a record for greatest length of wellbore drilled as a single connected system.

The project would have to be executed at very high temperature – much higher than encountered in typical oil and gas drilling environments. The 'high geothermal gradient' scenario from Beckers and Johnston (2022) (60°C/km) implies a temperature of 460°C (860°F) at reservoir temperature. Few geothermal wells have been drilled to these temperatures, and they faced a variety of complex challenges (Kruszewski and Wittig, 2018). Drilling 72 km of multilaterals at these temperatures would make the challenges exponentially greater.

Even in the 'medium geothermal gradient' scenario (30°C/km), 235°C (455°F), the downhole temperature would pose significant challenges for downhole tools. For context, the oil and gas industry defines an HPHT well (high pressure, high temperature) as one that exceeds 150°C (302°F).

**The #1 technical barrier is that this design requires dramatic reduction in the cost of drilling, while simultaneously executing exceptionally complex drilling plans at very high temperature.**

In comparison, conventional geothermal or EGS wells drilled to 460°C could hypothetically produce 10x more electricity than a closed-loop design while requiring 10x less drilling (Petty, 2022). This is because heat transport from *forced convection* is much more rapid than heat transport through conduction.

**The #2 technical barrier is that this design requires perfect sealing of ~72 km of uncased and uncemented wellbore, even as the wellbore and surrounding formation cool over time.**

The design requires a sealing material along the exterior of the uncased/uncemented laterals. If there is any imperfection in the sealant anywhere along the ~72 km in the laterals, then the working fluid (which is not water in many designs and may be expensive and environmentally sensitive) will leak off into the reservoir. Alternatively, fluid might flow into the wellbore from the surrounding reservoir. Either way, fluid exchange in or out of the closed-loop would hurt the efficiency of the system. Even if the seal is perfect *initially*, as the wellbore cools down and the surrounding rock (and sealant) contracts, there will be a tendency for cracks to form (Tarasovs and Ghassemi, 2010), breaking through the sealant. This is especially an issue because deep, hot geothermal wells tend to be drilled in formations with low ductility.

Many references have investigated improving production by optimizing the well heat transfer coefficient or the properties of the working fluid. These optimizations can improve production, but only to a point – the main 'limiting factor' for production is heat conduction inward from the surrounding rock.

Interestingly, because heat conduction through the formation is the primary limiting factor, simple analytic expressions for heat conduction yield reasonably accurate 'order of magnitude' estimates for the performance of real systems (McClure, 2021; Fowler and McClure, 2021; van Wees, 2021).

Malek et al. (2022) performed a techno-economic analysis of a deep multilateral closed-loop heat exchanger system. For their baseline scenario, they estimated a specific capital cost of $498,000-785,000/kWe installed capacity. With more aggressive assumptions using "state-of-the-art drilling technologies," they estimated $145,000/kWe. Assuming "significant advances of the current state-of-the-art," they estimated these costs could reach $37,000/kWe. They report that "the $2000-5000/kWe range [is] typically considered economical."

Providing a much more optimistic perspective, Figure 5.10 from Schulz and Livescu (2023) estimates that the cost of producing electricity from a deep closed-loop system in Texas using 2020 technology would be 7-22 cents per kW-hr, with a median case of 15 cents per kW-hr. From their citation, it appears that this estimate is based on the most aggressive, hypothetical drilling cost assumptions used in the Beckers and Johnston (2022) sensitivity analysis.

Let's consider the cost assumptions needed to achieve a levelized cost of 15 cents per kW-hr in Texas. Geothermal gradients in the hottest parts of Texas (Blackwell et al., 2011) are roughly comparable to the medium gradient (30°C/km) scenario from Beckers and Johnston (2022). In this scenario, their projected energy production was 2.2 MWe. At 15 cents per kW-hr, 2.2 MWe would generate $2.9M of revenue per year. That revenue would have to cover the cost of drilling 90 km of wellbore with 12 multilaterals at ~235°C, plus the cost of the power plant, the working fluid, the cost of capital, and other expenses.

Beckers and Johnston (2022) and White et al. (2023) performed techno-economic analyses of conduction-dominated closed-loop designs under a wide range of assumptions. With their most optimistic assumptions, they reach levelized cost estimates as low as 7 cents/kW-hr (Beckers and Johnston, 2022) and 8.3 cents/kW-hr (White et al., 2023). However, these are hypothetical scenarios considered as part of a sensitivity analysis, using extremely optimistic assumptions about thermal gradient, formation thermal conductivity, and drilling cost. They are not presented as projections of actual performance.

If large improvements in drilling technology did occur in the future, it still may not be optimal to develop deep closed-loop heat exchangers. Instead, extremely low cost, deep, high temperature, and extended reach wells could produce significantly more power (at lower levelized cost) if utilized for conventional geothermal wells or EGS.

*Thermally conductive fractures*

Another proposal is to create hydraulic fractures around the well and fill them with thermally conductive material (Ahmadi and Dahi-Taleghani, 2017; https://www.xgsenergy.com/). Heat will conduct into the thermally conductive fractures from the surrounding formation, and then conduct down the fractures towards the well, where there is a closed-loop heat exchanger.

This concept was analyzed by Fowler and McClure (2021). The analysis showed that for any realistic set of modeling assumptions, thermally conductive fractures provide little contribution to the total energy production of the well. To achieve significant energy contribution from the conductive fractures, they would need to have a thermal conductivity at least an order of magnitude greater than any known substance.

Why don't the thermally conductive fractures make a greater contribution? Because the rate of heat conduction is proportional to cross-sectional area, and the cross-sectional area is proportional to aperture. Crack apertures are very small, and so, the capacity for heat conduction along cracks is very low.

Is it possible to fill the cracks with a material so thermally conductive that they transport significant heat despite low aperture?

The effective 'draining length' of a crack (for either fluid flow or thermal conduction) is related to: (a) the capacity for transport *into* the crack from the surrounding formation, and (b) the capacity for transport *along* the crack towards the well (https://www.resfrac.com/blog/a-new-approach-for-interference-test-analysis-quantifying-the-degree-of-production-impact). Rock is about 1000x less thermally conductive than the most thermally conductive materials known. In contrast, when we create a propped hydraulic fracture in a very tight formation, the permeability of the crack (its capacity for fluid flow) is roughly 100,000,000-1,000,000,000x greater than the permeability of the surrounding rock. Thus, the effective draining length for fluid flow is 100s of feet, but the effective 'draining' length for thermal conduction is much lower.

Can we increase the number or thickness of the fractures?

Let's make a simple order of magnitude estimate. The Young's modulus of granite is ballpark 6 million psi. The total amount of 'net pressure' that you can build up when performing hydraulic fracturing – even with a thick viscous fluid – is around 1000 psi. Thus, the total amount of 'strain' that you can put on rock in the direction of Shmin during fracturing is ballpark 1000/6,000,000 = 1.67e-4. If you drill a 10,000 ft lateral, then a strain of 1.67e-4 implies a total displacement of 1.67 ft, or 0.7 m. That 0.7 m of displacement could be spread across 1000 fractures with aperture of 0.7 mm or 10,000 fractures with aperture of 0.07 mm. Either way, it represents an approximate 'upper bound' on how much thermally conductive material can be placed along the wellbore, due to physical limitations of the stiffness and strength of the rock.

In Fowler and McClure (2021), we assumed 400 fractures with aperture of 2.5 mm – which works out to 1 m of displacement, similar to the quantity in the back-of-the-envelope value above.

It doesn't matter whether there are fewer thicker cracks or more numerous thinner cracks. For the same total thickness of the cracks along the well, the system will have roughly similar performance. This is because thermal conduction *along* the cracks (not thermal conduction *into* the cracks) is the limiting factor for this system, and the capacity for conduction *along* the cracks scales with the aggregate thickness of all the cracks along the wellbore.

**The key technical barrier for this design is that it requires materials that are significantly more conductive than any known material. The quantity and thickness of the cracks cannot be increased beyond a certain point because they are inherently limited by rock strength and stiffness.**

In a preliminary study on thermally conductive fractures, Ahmadi and Dahi-Taleghani (2017) arrived at much more optimistic projections for power production. However, the calculations implicitly assumed that the fractures have infinite thermal conductivity.

*Natural convection*

Some closed-loop proposals use free convection to bring energy into the wellbore. The idea is to leave an openhole annulus outside the closed-loop heat exchanger. In the outer annulus, fluid circulates into the top of the annular region, cooling as it exchanges energy with the inner annulus (which is the outer section of the heat exchanger), and then flows out of the bottom of the annular region back into the formation. Circulation is driven by the density difference between the cooler fluid in the outer annulus and the hotter fluid in the reservoir. In the heat exchanger, fluid flows down the outer pipe, and back up the inner pipe.

Two proposed designs using natural convection are: (a) 'heat root' (https://blog.metamaterial.com/sage-meta-unlocking-the-potential-of-geothermal-for-clean-compact-renewable-baseload-power), and (b) the design from Higgins et al. (2021).

Let's start with the 'heat root' design. The idea is to drill a vertical well, and then form a hydraulic fracture from that well that propagates downward into hotter rock. Then, a downhole heat exchanger is installed within the well. Free convection circulates fluid in the hydraulic fracture, bringing energy to the production well from the underlying formation.

Propagating a hydraulic fracture downward 1000s of ft would be technically challenging. However, it may be possible, at least in some geologic settings, if using specialized fracturing fluids.

**The #1 technical barrier for this design is that it relies on flow through just one fracture (ie, stimulation from a single stage in a vertical well).**

In EGS, it has long been recognized that economic performance requires flow through a large number of fractures (Tester et al., 2006). The flow conductivity of a single fracture will be insufficient to sustain the high flow rates needed for economic viability, nor will it access sufficient surface area to achieve long-term sustainability. This is why current EGS projects are pursuing multistage stimulation designed to create 100s or 1000s of fractures, and they are experimenting with proppant to improve the flow rate achievable through those fractures.

**The #2 technical barrier is that free convection of liquid water generates only weak pressure gradient to drive flow.**

If we assume that a fracture is created with height of 1 km (which is an extremely optimistic assumption), and a 'high' geothermal gradient of 60˚C/km, then the density difference between the fluid at the top and bottom of the fracture will be ballpark 60 kg/m^3.

The most optimistic assumption would be to assume that the density difference is applied over the entire fracture height – 60 kg/m^3 over 1 km would be 0.6 MPa (86 psi) to drive circulation. But this estimate is an upper bound. As hot water rises from the underlying part of the fracture (or as cool water sinks downward from the heat exchanger), it will equilibrate in temperature with the surrounding rock. This will prevent the high temperatures at the bottom of the fracture from reaching the top of the fracture, and vice-versa. This will cause the density/temperature difference available to drive free convection to be much lower than this upper bound of 0.6 MPa – by at least 1-2 orders of magnitude.

By comparison, circulation between geothermal wells is typically driven by 3-20 MPa of pressure difference between the BHP of the injection and production wells.

A recent 'open' proposal by Polpis Systems (https://www.thinkgeoenergy.com/polpis-systems-publishes-open-proposal-for-supercritical-co2-geothermal/) is similar to the heat root design but includes thermally conductive proppant and circulates CO2 instead of water. This design has similar challenges as the heat root design. As discussed above, increasing the thermal conductivity of the proppant will have negligible impact on the performance of the system. Using CO2 instead of water will increase the deltaP created by density difference to drive free convection (because the density of CO2 is more sensitive to temperature than water), but overall, the use of CO2 would only modestly mitigate the problems faced by the design. Also, it is unclear how the CO2 in the fracture would be prevented from mixing with the fluid in the surrounding formation.

The second notable 'free convection' closed-loop design is from Higgins et al. (2021).

The design relies on steam entering the top of the annular region from an upper feed zone. The steam condenses into water as it flows down the annulus, and then it flows out a lower feed zone at the bottom of the annular region. A schematic of the design is shown in Figure 1 from Higgins et al. (2021), accessible at this link: <https://www.geothermal-library.org/index.php?mode=pubs&action=view&record=1034364>.

The density difference between steam and liquid water is large, and so in comparison with the 'heat root' design discussed above, this design will find it easier to generate pressure gradient to drive flow from free convection. At 260˚C, the saturation pressure of water is 4.69 MPa (680 psi). The density of liquid water at these conditions is 783 kg/m^3, and the density of steam is 23.7 kg/m^3. Over 1000 ft, the density difference would generate 2.3 MPa (334 psi). This is lower than the pressure difference that could be generated by opening the wellbore to flow at the surface, but at least, it is within the ballpark.

**The #1 technical barrier is that this design requires a very specific set of geologic conditions – sufficiently low pressure that steam is present in the reservoir, sufficient natural permeability to enable high flow rates into the well, and at least two major feed zones, which must have significant separation vertically.**

These requirements describe the conditions present only in a subset of wells in a subset of conventional geothermal reservoirs. This limits the applicability of the concept because the limiting factor for growth in geothermal energy production globally has been that conventional geothermal reservoirs are relatively scarce.

**The #2 technical barrier is that it's unclear whether this design would ever be preferred to simply producing the well to the surface, which is what's done in a conventional geothermal well.**

In a conventional geothermal well, fluid is produced to the surface. The expansion of the fluid as it flows up 1000s of ft of wellbore reduces density, reduces hydrostatic head, and helps suck fluid inwards from the reservoir.

I have not done a detailed analysis, but an initial analysis suggests that in the great majority of practical scenarios (possibly, in all practical scenarios), the design from Higgins et al. (2021) will generate significantly *less* hydrostatic head difference to drive flow into the well, relative to the conventional design where fluid simply flows to the surface. Also, the design has a relatively inefficient heat exchanger because the design is co-current, rather than countercurrent.

Are there any conditions when the design from Higgins et al. (2021) would produce energy into the well at a higher rate than a conventional geothermal well? I am not sure; it is possible. I have not yet found a paper in the literature that addresses this question.

Higgins et al. (2021) argue that the design is advantageous because it does not remove fluid from the reservoir, which maintains reservoir pressure. But this objective is already accomplished routinely with conventional geothermal approaches – using a *surface* heat exchanger followed by reinjection. With the widely-used binary cycle power plant design, fluid is produced to the surface, put through a heat exchanger, and then reinjected.

Relative to placing the heat exchanger downhole, a surface heat exchanger has significant advantages: (a) the energy transport rate into the production wells is maximized by flowing fluid directly to the surface, (b) the heat exchanger is much simpler and easier to maintain, and (c) the heat exchanger can be designed to operate more efficiently.

Finally, we should note a misconception about this design, which I have seen repeatedly expressed in the literature. It is sometimes stated that a field trial of the Higgins et al. (2021) design was performed at Coso geothermal field, and it produced 1.2 MWe. However, the field trial did not actually test the closed-loop design described by Higgins et al. (2021). As explained in an earlier paper by Higgins et al. (2019), the Coso field trial was performed while placing the heat exchanger inside of a wellbore *that was producing fluid from the reservoir to the surface, as in a conventional geothermal well*. This is not the same thing as placing a closed-loop heat exchanger inside a shut-in well, relying on free convection, which is the Higgins (2021) design.

So then, if the field trial from Higgins et al. (2019) produced 1.2 MWe, why not use that design, and place downhole heat exchangers in geothermal wells as they produce fluid to the surface? Because there is not an engineering reason to do so. Placing a downhole heat exchanger inside a well that is producing to the surface will *reduce* the energy production of the well, relative to the conventional approach where the fluid is simply produced to the surface. The heat exchanger will cool the fluid in the well – increasing hydrostatic head – and increase frictional resistance to flow. Further, as noted above, the downhole heat exchanger will be inefficient because it is co-current and not countercurrent.

Thus, in the 1.2 MWe field pilot described by Higgins et all. (2019), the system would have produced *more* power if there had *not* been a heat exchanger in the well, and if it had instead been produced as a conventional geothermal well. My understanding is that the configuration was used to facilitate the testing of downhole equipment, rather than as a full-scale commercial test.

*Retrofitting old oil and gas wells*

Numerous authors have suggested that we could install closed-loop heat exchangers in depleted oil and gas wells. The advantage would be that the well has already been drilled, reducing cost.

Some disadvantages are:

1. The cost is not actually *that* low. Insulated tubing is quite expensive, and the wellbore sealing operations could be complex.

2. It would be technically challenging to achieve a perfect seal between the wellbore from the surrounding formation. For example, a typical shale well has 1000+ perforation holes along its lateral, many of which erode significantly during fracturing. These would all need to be sealed. At the same time, the installation of a heat exchanger would require that the interior of the well must remain open to flow, so that insulated tubing can be run into the hole, and fluid circulated down and back up.

3. The majority of oil and gas wells are not drilled to very high temperature, reducing their geothermal energy production potential. Most hydrocarbon reservoirs are 70-130°C, low by geothermal standards.

**4. Above all, the most important technical barrier is that conduction-based closed-loop heat exchangers produce little energy per ft of wellbore, and so the energy production from retrofitted oil and gas wells will be very low.**

As noted above, Beckers and Johnston (2022) analyzed the design from Toews and Holmes (2021) and calculated that a conduction-dominated closed-loop heat exchanger design at 30°C/km and 7.5 km depth (235°C at reservoir depth), and 12 laterals at a total wellbore length of 90 km will produce 2.2 MWe.

A typical shale well has 6 km of wellbore (3 km of lateral) at ballpark 120°C. So, with 24x less lateral length, 15x less total wellbore length, and 125°C lower reservoir temperature, how much energy production should we expect? A back-of-the-envelope calculation suggests that it produce 2-3 orders of magnitude less power. Let's say hypothetically that a retrofitted shale well could produce 20 kWe. At the optimistic assumption of 10 cents per kW-hr, that's $17,500/year. It's hard to see how that pays for 20,000 ft of insulated tubing, a treatment to seal the 1000+ perforations in the casing along the lateral, surface facilities, and the cost of capital.

In a preliminary study on this topic, Livescu and Dindoruk (2022) concluded that there is "great potential" from retrofitting existing oil and gas well with downhole heat exchangers. However, in the calculations, they neglected thermal drawdown in the formation (equivalent to assuming that the formation has infinite thermal conductivity). Consequently, they arrived at estimates for power production per well orders of magnitude higher than would be achieved in practice. With more realistic assumptions, the idea looks less promising.

*Producing heat instead of electricity*

It has been suggested that the economics of closed-loop geothermal could be improved by targeting direct-use of heat, instead of electricity generation (Beckers and Johnston, 2022; White et al., 2023). Let's consider why heat production may provide better economics than electricity production, and then consider why it may not.

Figure 13 from Zarrouk and Moon (2014) shows the efficiency of actual geothermal power plants. Efficiency is a function of temperature/enthalpy and ranges from 1-2% (at temperatures from 100˚C and below) to 17% (at enthalpy as high as 2850 kJ/kg). Efficiency can be defined in different ways; these efficiency numbers appear relatively low because they are reported relative to a reference temperature of 0˚C. Also, their reported efficiency values incorporate all parasitic loads in the plant, and other practical factors, such as design considerations needed to prevent scaling of dissolved solids.

As an example, let's consider a conventional geothermal well producing at 200˚C. From Zarrouk and Moon (2014), the power plant efficiency will be approximately 7.5%. Thus, a well producing 50 kg/s will be producing 42.5 MWth (relative to a reference temperature of 0˚C), which could be converted to ~3.2 MWe. Typical wholesale electricity prices are in the ballpark of 3-5 cents per kW-hr. White et al. (2023) estimates that typical industrial heating cost is 2.65 cents per kW-hr. Given this comparison – why would anyone ever generate electricity from a geothermal well? Why would we build costly power plants to perform a conversion at 7.5% efficiency from something worth 2.65 cents per kW-hr into something worth 3-5 cents per kW-hr? At face value, the heat appears to be 7-11x more valuable than the electricity that can be generated from it.

The reason is that there is a limited *market* for geothermal heat. For electricity, there are massive existing distribution systems and widespread demand. But to utilize heat from warm water produced from a geothermal well, it is impractical to transport heat long distances and specialized equipment and pipelines are needed. For example, geothermal district heating and cooling requires construction of pipelines to distribute the heat between buildings.

Thus, when we consider geothermal direct-use, the most important limiting factor is not the cost of producing the heat, but rather, the cost and practicality of building surface facilities designed to utilize the heat.

I am 100% in favor of building more infrastructure and industrial applications to increase utilization of geothermal direct-use. In fact, I expect this to happen in the coming years. But as this market grows, will it be optimal to use closed-loop heat exchangers to provide the heat?

Let's consider the alternative – directly producing water from the ground, as in a conventional geothermal well. There are many natural geologic formations that: (a) contain warm water (>70-120˚C), and (b) have high permeability (ie, high natural capacity for fluid flow).

When we talk about the 'limiting factors' for geothermal growth, we often talk about the scarcity of formations with high temperature and high permeability. But in this context, we're talking about formations *hot enough to produce electricity at reasonable efficiency*, which is generally viewed as being around 120-150˚C or higher. If we consider *direct-use*, then the minimum temperature threshold is much lower, which means that a much wider range of geologic formations are available. Naturally permeable saline aquifers are abundant in sedimentary formations around the world. Simple vertical wells can be used to cheaply produce large volumes of warm water from these formations.

To summarize, aside from the general issue of cost (all geothermal options would need to be reasonably cost-competitive with alternative sources of heat, such as natural gas), there are at least two major issues that would need to be addressed to reach widespread adoption of closed-loop heat exchangers for direct-use:

**1. There needs to be sufficient demand for heat** *in the form of warm fluid flowing from a downhole heat exchanger.*

**2. The closed-loop heat exchanger needs to produce heat at lower cost than would be achieved from drilling conventional wells to produce fluid from permeable subsurface formations.**

These comments also apply to the retrofitting of existing oil and gas wells. There are many existing oil and gas wells that produce warm fluid (>70-120˚C) to the surface at high water cut. This warm water is available at the surface *for free*; yet in the vast majority of cases, it is reinjected back into the subsurface without any energy extraction. This fluid could be exploited for geothermal direct-use (https://www.vox.com/recode/23024204/geothermal-energy-heat-oil-gas-wells).

Notwithstanding these comments, there is a good technical reason to conclude that direct-use is more advantageous for downhole closed-loop heat exchangers than electricity generation. Most closed-loop designs rely on heat conduction to bring fluid to the well; conduction is driven by the temperature difference between the well and the formation. As a result, in most closed-loop designs (unlike in conventional geothermal wells), the production temperature at the surface is much lower than the temperature in the reservoir (Beckers and Johnston, 2022; White et al., 2023). For example, in the Beckers and Johnston (2022) analysis, the system drilled to 235˚C reservoir temperature only produces fluid at a temperature of 120-140˚C (their Figure 7). Therefore, unless drilling into exceptionally hot formations, closed-loop heat exchangers will be producing fluid at relatively low enthalpy (by geothermal standards), and this will result in low efficiency for electricity generation.

*Recap*

Considering the discussion above, my assessment is that the 'best case scenario' for closed-loop geothermal would be: (a) a purely conductive closed-loop design, (b) a market with high willingness to pay for relatively low enthalpy fluid that can be utilized for direct-use (rather than for electricity generation), (c) a site that lacks naturally permeable formations in the stratigraphic column, so that production from conventional wells is not feasible, (d) large decreases in the cost of drilling, and (e) development of materials that can perfectly seal uncased/uncemented laterals and avoid developing cracks as the formation progressively cools down.

I am interested for feedback. Do you think I missed something important about the designs discussed above? Are there any other closed-loop designs that I missed?